\documentclass[aps,prd,reprint,superscriptaddress,nofootinbib,longbibliography]{revtex4-2}

\usepackage{graphicx}
\usepackage{amsmath,amssymb}
\usepackage{bm}
\usepackage{xcolor}
\usepackage[colorlinks=true,allcolors=blue!50!black]{hyperref}
\usepackage{orcidlink}

\newcommand{\kms}{\mathrm{km\,s^{-1}}}
\newcommand{\mss}{\mathrm{m\,s^{-2}}}
\newcommand{\Msun}{M_\odot}
\newcommand{\Lsun}{L_\odot}
\newcommand{\HI}{H\,{\sc i}}
\newcommand{\LCDM}{$\Lambda$CDM}
\newcommand{\Yd}{\Upsilon_{\rm disk}}
\newcommand{\Yb}{\Upsilon_{\rm bul}}
\newcommand{\dBIC}{\Delta{\rm BIC}}
\newcommand{\chin}{\chi^2_\nu}

\begin{document}

\title{Cusps, cores, and one acceleration:\\
dark-matter haloes versus modified dynamics across the full SPARC sample}

\author{Mikhael Da Silva\,\orcidlink{0009-0002-4029-9628}}
\affiliation{Independent Researcher}

\date{September 5, 2026}

\begin{abstract}
The rotation curves of the 175 disk galaxies in the SPARC database are confronted, one by one and under identical conditions, with the two mass models that have dominated the missing-mass debate for four decades: a cold dark-matter halo of Navarro--Frenk--White (NFW) form, and Milgromian dynamics (MOND) with the acceleration constant held at $a_0=1.2\times10^{-10}\,\mss$. Stellar mass-to-light ratios, distances and inclinations are treated as nuisance parameters with priors set by the $3.6\,\mu$m photometry and by the SPARC error budget, and every galaxy is sampled with a Markov-chain Monte Carlo. Three findings emerge. (i) Judged by the Bayesian information criterion, the free NFW halo is preferred in 96 galaxies and MOND in 79; once the halo is required to obey the \LCDM\ concentration--mass relation the score becomes 86 to 89, i.e.\ a draw. The two theories fail in different places: MOND loses in massive, high-surface-brightness, bulge-dominated systems, NFW loses in slowly rising dwarf rotation curves. (ii) The NFW concentrations demanded by the data sit a median $0.24$~dex below the \LCDM\ expectation, and 53\% of the sample lies more than $2\sigma$ below it, independent of halo mass; this is the core--cusp problem expressed as a scaling-relation offset rather than as a handful of anecdotes. (iii) The radial acceleration relation built from the full sample has $a_0=(1.13\pm0.02)\times10^{-10}\,\mss$ and an observed scatter of $0.14$~dex, yet the value of $a_0$ preferred by individual galaxies spans more than a decade and correlates with surface brightness. The two statements are not in contradiction: the population-level relation is tight because it averages over exactly the degeneracies with mass-to-light ratio and distance that make $a_0$ ill-determined galaxy by galaxy. The full table of fits is provided.
\end{abstract}

\maketitle

\section{Introduction}
\label{sec:intro}

Ninety years after Zwicky noted that the Coma cluster moves too fast for its light \cite{Zwicky1933}, and fifty after the flat rotation curves of spiral galaxies made the same point in a form nobody could dismiss \cite{RubinFord1970,Rubin1980,Bosma1981}, the missing-mass problem still admits two live explanations that are conceptually as far apart as physics allows. Either galaxies are embedded in haloes of a collisionless, non-baryonic substance that outweighs the visible matter by an order of magnitude, or the law of gravity (or inertia) departs from Newton's at accelerations below $a_0\approx10^{-10}\,\mss$ \cite{Milgrom1983,Milgrom1983b}. The first option, cold dark matter (CDM), is embedded in a cosmological model of spectacular success on scales larger than a megaparsec \cite{Planck2018}. The second, MOND, has for decades been embarrassingly good at predicting the internal dynamics of individual galaxies from their baryons alone \cite{Begeman1991,SandersMcGaugh2002,FamaeyMcGaugh2012}, and has no accepted relativistic completion that survives all cosmological tests.

It is worth being clear about what a rotation-curve study can and cannot settle. It cannot decide between the two paradigms; the cosmic microwave background, the growth of structure, gravitational lensing by clusters and the bullet-cluster geometry all weigh in, and they mostly weigh against modified dynamics. What rotation curves can do, and what nothing else does as sharply, is test each hypothesis on the ground where it makes its most specific predictions. \LCDM\ predicts a particular family of halo density profiles \cite{NFW1997} with a particular concentration--mass relation \cite{DuttonMaccio2014}; MOND predicts the rotation curve outright, given the baryons and one universal constant. Both predictions are falsifiable at the level of individual galaxies, and both have been under pressure for years: CDM from the core--cusp and diversity problems \cite{deBlok2010,Oman2015,BullockBK2017}, MOND from claims that $a_0$ is not, in fact, universal \cite{Rodrigues2018}, claims that were immediately contested \cite{McGaugh2018,Kroupa2018}.

The Spitzer Photometry and Accurate Rotation Curves (SPARC) database \cite{Lelli2016} is the natural arena for this contest. It provides 175 galaxies with \HI/H$\alpha$ rotation curves and $3.6\,\mu$m surface photometry, spanning four decades in luminosity and three in surface brightness, all reduced with a homogeneous procedure. It is the sample in which the radial acceleration relation (RAR) was discovered \cite{McGaugh2016,Lelli2017}, and in which the most extensive catalogues of halo fits have been assembled \cite{Katz2017,Li2020}. Much of the literature, however, examines one side at a time: halo models with cosmological priors on one hand \cite{Katz2017,Li2020}, the RAR or MOND on the other \cite{Rodrigues2018,Li2018,Marra2020}. Comparisons across the divide are scarcer, and are often made with different nuisance treatments, different subsamples, and different statistics, which leaves the reader to reconcile them.

This paper does the reconciliation explicitly. The same 175 galaxies are fitted with an NFW halo, with an NFW halo constrained by the \LCDM\ concentration--mass relation, and with MOND, using the same likelihood, the same priors on the stellar mass-to-light ratio, distance and inclination, the same sampler, and the same model-comparison statistic. No galaxy is excluded, no rotation-curve point is edited, and no parameter is tuned to make either side look good. The aim is not to declare a winner---the data, as will be seen, decline to do so---but to characterise \emph{where} and \emph{why} each description strains, and to put on the same page two results that are usually kept apart: the tightness of the population-level acceleration relation, and the looseness of the per-galaxy acceleration scale.

The paper is organised as follows. Section~\ref{sec:data} describes the data. Section~\ref{sec:models} sets out the baryonic mass model and the two dynamical hypotheses. Section~\ref{sec:method} presents the likelihood, priors, sampling and model-comparison strategy. Section~\ref{sec:results} reports the results, Section~\ref{sec:discussion} discusses what they do and do not imply, and Section~\ref{sec:conclusions} concludes. Appendix~\ref{app:table} lists the fit for every galaxy.

\section{Data}
\label{sec:data}

SPARC \cite{Lelli2016} collects 175 late-type galaxies (Hubble types S0 to Im and BCD) for which extended \HI\ rotation curves, occasionally supplemented by H$\alpha$ data in the inner regions, exist alongside Spitzer $3.6\,\mu$m imaging. For each galaxy the database supplies the observed rotation velocity $V_{\rm obs}(R)$ with its uncertainty $\delta V_{\rm obs}$, and the Newtonian circular-velocity contributions of the gas disk, stellar disk and bulge, $V_{\rm gas}$, $V_{\rm disk}$ and $V_{\rm bul}$, obtained by solving Poisson's equation for the observed surface density profiles under the assumption of a stellar mass-to-light ratio of $1\,\Msun/\Lsun$ at $3.6\,\mu$m. The gas contribution includes the standard correction for helium; where the gas surface density has a central depression, $V_{\rm gas}$ is reported as negative, signifying an outward force, and this sign convention is retained here.

\begin{figure}
\includegraphics[width=\columnwidth]{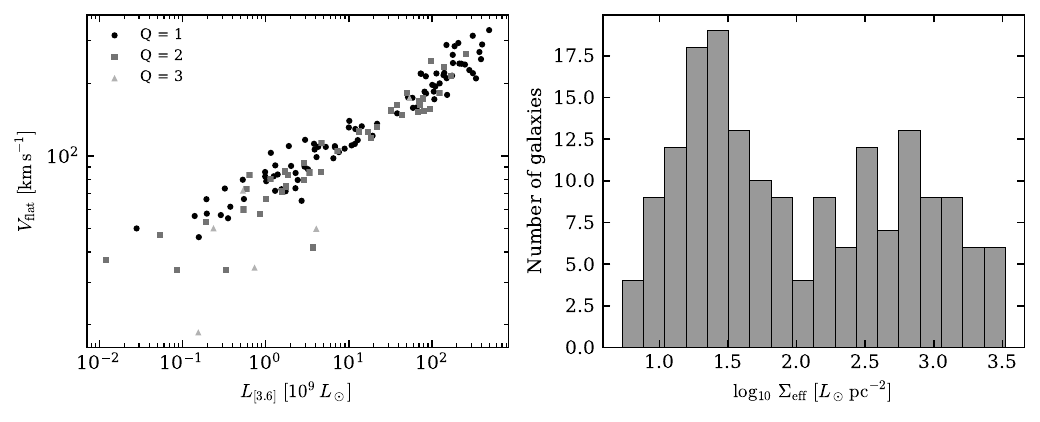}
\caption{The SPARC sample. Left: flat rotation velocity against $3.6\,\mu$m luminosity for the 135 galaxies whose curves reach a flat part, coded by quality flag. Right: distribution of effective surface brightness for all 175 galaxies. The sample is not volume-complete but covers the full range of disk-galaxy properties, which is what a comparison of dynamical laws requires.}
\label{fig:sample}
\end{figure}

The master table lists for each galaxy the adopted distance $D$ and its uncertainty, the inclination $i$ and its uncertainty, the total $3.6\,\mu$m luminosity $L_{[3.6]}$, the effective radius and surface brightness, the \HI\ mass, the asymptotically flat velocity $V_{\rm flat}$ where the curve reaches one, and a quality flag $Q\in\{1,2,3\}$ reflecting the reliability of the rotation curve (99, 64 and 12 galaxies respectively). Distances come from Hubble flow ($H_0=73\,\kms\,{\rm Mpc}^{-1}$, with a Virgocentric infall correction), tip of the red giant branch, Cepheids, supernovae, or membership of the Ursa Major cluster, with correspondingly different uncertainties; Hubble-flow distances typically carry 20--30\% errors, which will matter below. The sample spans $L_{[3.6]}$ from $10^7$ to $3\times10^{11}\,\Lsun$ and effective surface brightness from $\sim\!5$ to $\sim\!5000\,\Lsun\,{\rm pc}^{-2}$ (Fig.~\ref{fig:sample}). In total there are 3391 rotation-curve points, with a median of 14 per galaxy.

Nothing is removed from the sample. Low-quality ($Q=3$) galaxies and face-on systems are kept, because a test that discards its own most awkward cases is not a test; their influence is instead tracked by reporting statistics for the $Q=1$ subset alongside the full sample. Where a rotation curve has fewer points than the number of model parameters, the reduced $\chi^2$ is defined with one degree of freedom and the galaxy is retained in counts but carries negligible weight in any conclusion.

\section{Mass models}
\label{sec:models}

\subsection{Baryons}
\label{sec:baryons}

The baryonic contribution to the circular velocity is
\begin{equation}
V_{\rm bar}^2(R)=\Yd\,V_{\rm disk}^2+\Yb\,V_{\rm bul}^2+V_{\rm gas}\,|V_{\rm gas}|,
\label{eq:vbar}
\end{equation}
where $\Yd$ and $\Yb$ are the disk and bulge mass-to-light ratios at $3.6\,\mu$m in solar units and the last term preserves the sign convention of the gas term. Stellar-population synthesis at $3.6\,\mu$m predicts a mass-to-light ratio that is nearly independent of colour and star-formation history, with $\Yd\simeq0.5$ for disks and $\Yb\simeq0.7$ for the older, more metal-rich populations of bulges \cite{Schombert2014,Meidt2014}. These values are adopted as the centres of lognormal priors of width $0.1$~dex, the width recommended by \citet{Li2018} and \citet{Li2020}. This is the single most consequential assumption in any rotation-curve analysis; it is made explicit here so that its consequences can be traced in the results.

Distance and inclination enter the comparison in a way that is easy to state and easy to forget. If the true distance is $D=d\,D_0$, the radii scale as $R\to dR$ and every baryonic velocity as $V^2\to d\,V^2$, because the enclosed mass inferred from a surface-brightness profile scales with $D^2$ while the radius scales with $D$. If the true inclination is $i$ rather than the catalogued $i_0$, the observed rotation velocity becomes $V_{\rm obs}\,\sin i_0/\sin i$, with the error scaled identically. Both corrections are applied to the model rather than to the data, so that the residuals are always computed against the published velocities.

\subsection{Cold dark matter: the NFW halo}
\label{sec:nfw}

The dark matter halo is taken to have the density profile that emerges from collisionless $N$-body simulations \cite{NFW1997},
\begin{equation}
\rho(r)=\frac{\rho_s}{(r/r_s)(1+r/r_s)^2},
\label{eq:nfw}
\end{equation}
whose circular velocity is
\begin{equation}
V_{\rm NFW}^2(R)=V_{200}^2\,\frac{\ln(1+cx)-cx/(1+cx)}{x\,[\ln(1+c)-c/(1+c)]},
\label{eq:vnfw}
\end{equation}
with $x=R/R_{200}$, $R_{200}=V_{200}/(10\,H_0)$ and $M_{200}=V_{200}^3/(10\,GH_0)$ the radius and mass at which the mean enclosed density is 200 times the critical density, and $c=R_{200}/r_s$ the concentration. Two parameters, $(\log_{10}c,\,\log_{10}V_{200})$, thus describe the halo, and the total model velocity is $V^2=V_{\rm bar}^2+V_{\rm NFW}^2$.

Two versions of the halo hypothesis are considered. In the \emph{free} version the concentration and virial velocity are given flat priors in the logarithm, $0\le\log_{10}c\le2$ and $1\le\log_{10}(V_{200}/\kms)\le2.9$. This is a phenomenological halo that happens to have the NFW shape; it is the most generous thing one can do for cold dark matter. In the \emph{\LCDM} version the concentration is additionally required to follow the relation calibrated by \citet{DuttonMaccio2014} on Planck-cosmology simulations at $z=0$,
\begin{equation}
\log_{10}c_{200}=0.905-0.101\,\log_{10}\!\left(\frac{M_{200}}{10^{12}\,h^{-1}\,\Msun}\right),
\label{eq:cM}
\end{equation}
with a lognormal scatter of $0.11$~dex, and $h=0.671$. This is the halo that the cosmological model actually predicts, before baryonic feedback is invoked to reshape it. The difference between the two versions is therefore a measurement of how far the data ask the haloes to deviate from cosmological expectations.

The NFW profile is not the last word on CDM haloes. Feedback-modified profiles such as that of \citet{DiCintio2014}, or the cored profiles produced by self-interacting dark matter \cite{Ren2019}, fit rotation curves better \cite{Katz2017,Li2020}, and deliberately so. They are not used here because the aim is to test the prediction of the simplest, best-motivated version of each theory, not the most flexible. A model that is given more freedom will always fit better; the question is what the added freedom is buying and whether the base theory needs it.

\subsection{Modified dynamics: MOND}
\label{sec:mond}

In MOND \cite{Milgrom1983}, the observed centripetal acceleration $g_{\rm obs}=V_{\rm obs}^2/R$ is related to the Newtonian acceleration generated by the baryons, $g_{\rm bar}=V_{\rm bar}^2/R$, through
\begin{equation}
g_{\rm obs}=\nu\!\left(\frac{g_{\rm bar}}{a_0}\right)g_{\rm bar},
\label{eq:mond}
\end{equation}
where $\nu(y)\to1$ for $y\gg1$ and $\nu(y)\to y^{-1/2}$ for $y\ll1$. The interpolation function is not specified by the theory. The ``simple'' form \cite{FamaeyBinney2005},
\begin{equation}
\nu(y)=\tfrac{1}{2}+\sqrt{\tfrac{1}{4}+\tfrac{1}{y}},
\label{eq:simple}
\end{equation}
is adopted for the individual fits, and the empirical form of \citet{McGaugh2016}, $\nu(y)=[1-\exp(-\sqrt{y})]^{-1}$, is used alongside it when fitting the population-level acceleration relation. The two differ by at most a few per cent in the transition region and give indistinguishable results below. The acceleration constant is held fixed at $a_0=1.2\times10^{-10}\,\mss$ \cite{Begeman1991}, the value that has been in use since the 1990s; Section~\ref{sec:a0} revisits whether the data agree with it.

Equation~(\ref{eq:mond}) is an algebraic relation between accelerations in the same spherically symmetric approximation under which the SPARC baryonic velocities were computed. It neglects the non-linearity of the field equation in the aquadratic-Lagrangian and quasi-linear formulations, which for thin disks amounts to corrections at the several-per-cent level, well below the systematic uncertainties of the data. The MOND model has no free parameters beyond the nuisance parameters; this is its point, and it is also why it has to be compared carefully.

\section{Method}
\label{sec:method}

\subsection{Likelihood and priors}
\label{sec:like}

For each galaxy the log-posterior is
\begin{equation}
\ln\mathcal{P}=-\tfrac{1}{2}\sum_{k=1}^{N}\frac{\left[V_{{\rm obs},k}\,\frac{\sin i_0}{\sin i}-V_{\rm mod}(dR_k;\,\bm{\theta})\right]^2}{\left(\delta V_{{\rm obs},k}\,\frac{\sin i_0}{\sin i}\right)^2}+\ln\pi(\bm{\theta}),
\label{eq:post}
\end{equation}
where $\bm{\theta}=(\log_{10}c,\log_{10}V_{200},\log_{10}\Yd,\log_{10}\Yb,d,i)$ for the NFW models and $\bm{\theta}=(\log_{10}\Yd,\log_{10}\Yb,d,i)$ for MOND. The prior $\pi$ is the product of the lognormal priors on $\Yd$ and $\Yb$, a Gaussian on $dD_0$ centred on $D_0$ with the SPARC distance uncertainty, a Gaussian on $i$ centred on $i_0$ with the SPARC inclination uncertainty, and, for the \LCDM\ halo, the Gaussian of Eq.~(\ref{eq:cM}) in $\log_{10}c$. The distance and inclination are additionally truncated at $\pm3\sigma$ of their catalogued uncertainties. This truncation deserves comment. Without it, a handful of galaxies with poorly known Hubble-flow distances allow the fit to relocate them by factors of two or three, which is not a statement about gravity but about the freedom of a Gaussian tail. With it, the fit is held to what the observers actually measured. For galaxies without a bulge $\Yb$ is retained in the sampling but pinned by its prior and not counted as a free parameter.

\subsection{Sampling}
\label{sec:sampling}

Each posterior is first maximised from a grid of nine starting points in the halo parameters (one for MOND) with a Nelder--Mead simplex, and the maximum is then used to initialise an affine-invariant ensemble sampler \cite{emcee} with 32 walkers run for 1200 steps, the first 400 being discarded. The maximum-a-posteriori (MAP) point is taken as the better of the simplex solution and the highest-posterior sample; parameter uncertainties are the 16th and 84th percentiles of the marginal posteriors. With 525 fits (175 galaxies $\times$ 3 models) and a median of 14 data points per galaxy, this is a modest computation, and it was checked on a subset that doubling the chain length changes no quoted number.

\subsection{Model comparison}
\label{sec:bic}

Two models with different numbers of parameters cannot be compared on $\chi^2$ alone. The Bayesian information criterion \cite{Schwarz1978},
\begin{equation}
{\rm BIC}=\chi^2_{\rm MAP}+k\ln N,
\label{eq:bic}
\end{equation}
where $k$ is the number of free parameters and $N$ the number of data points, penalises the two extra halo parameters by $2\ln N$, which for the median galaxy is $5.3$. The quantity reported throughout is
\begin{equation}
\dBIC={\rm BIC}_{\rm NFW}-{\rm BIC}_{\rm MOND},
\label{eq:dbic}
\end{equation}
so that positive values favour MOND and negative values favour the halo. The conventional reading \cite{KassRaftery1995} is that $|\dBIC|<2$ is not worth mentioning, 2--6 is positive evidence, 6--10 is strong, and beyond 10 is very strong. The BIC is an approximation to the Bayesian evidence that assumes a well-peaked posterior and a large sample; neither holds for a rotation curve with eight points. It is used here because it is transparent, because it is what the literature uses, and because the conclusions below do not depend on the choice of penalty: the Akaike criterion and the raw $\chi^2$ give the same qualitative picture, as shown in Section~\ref{sec:comparison}. A proper evidence computation with nested sampling would be the next step and is left for future work, with the remark that for the strong cases in either direction, where $|\dBIC|$ runs into the tens or hundreds, no reasonable Occam factor would reverse the verdict \cite{Trotta2008}.

\section{Results}
\label{sec:results}

\subsection{Overall fit quality}
\label{sec:quality}

\begin{figure}
\includegraphics[width=\columnwidth]{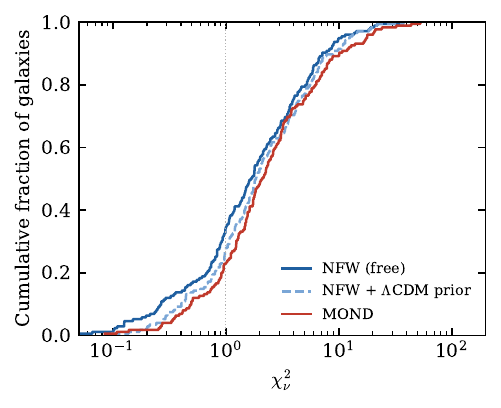}
\caption{Cumulative distribution of the reduced $\chi^2$ at the maximum-a-posteriori point for the three models, over all 175 galaxies. The dotted line marks $\chin=1$. The free NFW halo has the lowest values, the \LCDM-constrained halo pays a visible price for the concentration prior, and MOND, with two fewer parameters, sits slightly to the right; the three curves are nevertheless closer to one another than the four-decade-old debate might lead one to expect.}
\label{fig:chi2}
\end{figure}

\begin{figure*}
\includegraphics[width=\textwidth]{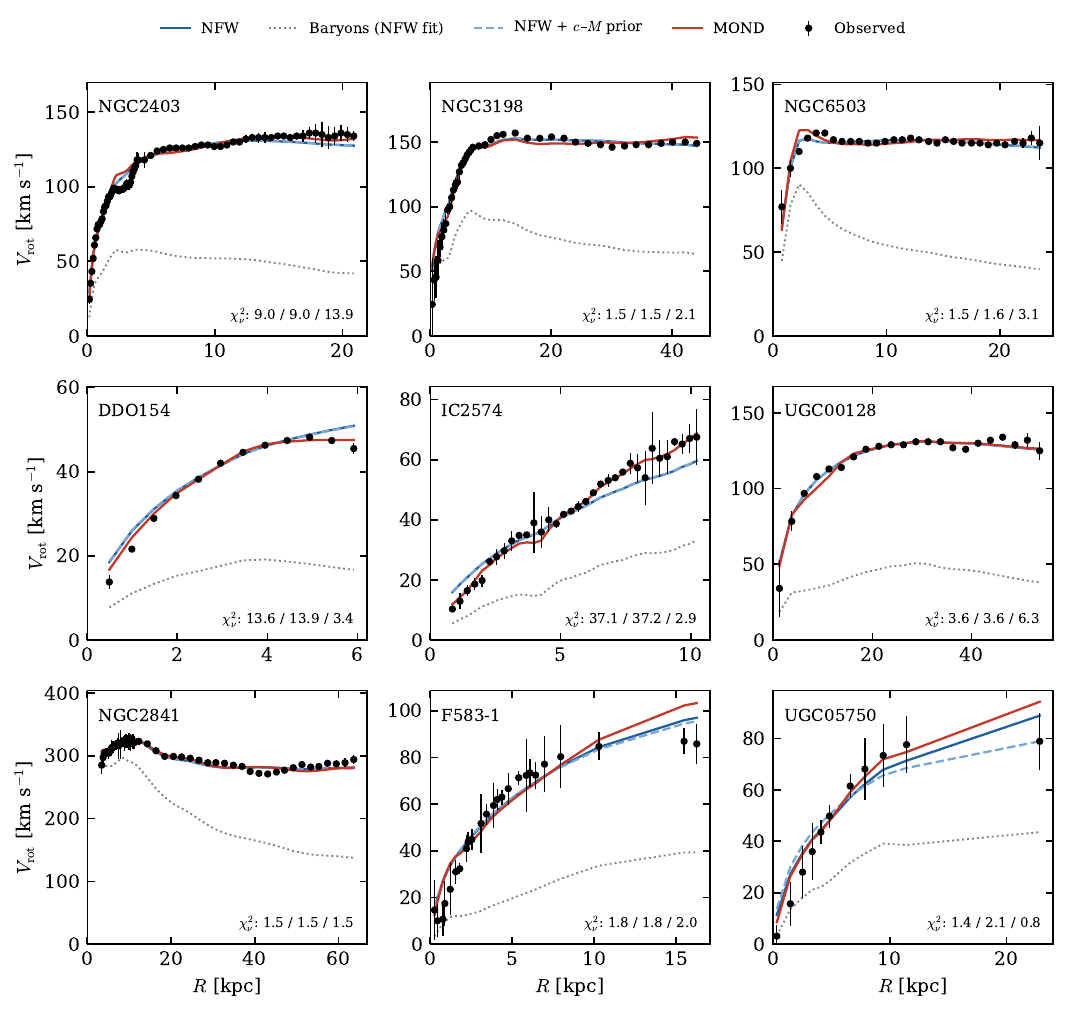}
\caption{Nine galaxies chosen to span the range of outcomes. Points are the observed rotation curves with their published errors; curves are the maximum-a-posteriori models, transformed back to the observed distance and inclination so that all three can be compared against the same data. The dotted grey curve is the baryonic contribution in the NFW fit. Numbers in each panel are $\chin$ for NFW, NFW+\LCDM\ prior, and MOND. Top row: three high-quality spirals in which both descriptions work but NFW works better (NGC~2403, NGC~3198, NGC~6503). Middle row: three gas-rich dwarfs and low-surface-brightness disks in which the slowly rising curve defeats the cusp (DDO~154, IC~2574, UGC~00128). Bottom row: a massive early-type spiral with a bulge (NGC~2841), and two low-surface-brightness galaxies with Hubble-flow distances where the verdict depends on how far the distance may be moved (F583-1, UGC~05750).}
\label{fig:examples}
\end{figure*}

Figure~\ref{fig:chi2} shows the cumulative distribution of $\chin$ for the three models, and Fig.~\ref{fig:examples} shows nine representative fits. The median reduced $\chi^2$ is $1.68$ for the free NFW halo, $1.85$ for the halo with the \LCDM\ prior, and $2.10$ for MOND; restricted to the 99 galaxies with $Q=1$ the medians are $1.48$, $1.56$ and $1.82$. The fraction of galaxies with $\chin<1$ is 34\%, 27\% and 23\%; the fraction with $\chin>5$ is 21\%, 23\% and 25\%. Summed over all 3391 points, the total $\chi^2$ is 9969, 10354 and 13514.

Three remarks. First, none of the models achieves $\chin\approx1$ for the bulk of the sample. This is not news: the published errors on $V_{\rm obs}$ describe the precision of the tilted-ring fit, not the accuracy of the assumption that gas moves on circular orbits in an axisymmetric potential, and non-circular motions, warps and beam smearing all contribute residuals that no smooth model can absorb \cite{Lelli2016,Li2018}. A $\chin$ of 2 is, for this data set, a good fit. Second, 77 galaxies are fitted with $\chin<2$ by both NFW and MOND, and 26 are fitted with $\chin>5$ by both. Almost half of the sample, in other words, does not discriminate between the two laws at all, either because both work or because neither does; the interesting physics lives in the remaining half. Third, the \LCDM\ prior costs the halo model a median $\Delta\chi^2=0.7$, which is negligible, but for nine galaxies the cost exceeds 10 and for one (IC~2574) it is meaningless because even the free halo fails. The prior, in other words, is mostly harmless and occasionally fatal, a pattern that recurs below.

\subsection{Model comparison}
\label{sec:comparison}

Table~\ref{tab:counts} summarises the comparison. With the free NFW halo the BIC prefers the halo in 96 galaxies and MOND in 79; with the \LCDM\ concentration prior, which removes most of the halo's freedom without removing any of its parameters, the count is 86 to 89. Restricting to $Q=1$ the halo does somewhat better (58 to 41, and 55 to 44 with the prior). The Akaike criterion, which penalises parameters less severely, shifts every count a little towards the halo. In none of these tallies does either model win a clear majority, and the median $\dBIC$ over the sample is $-0.9$ (free halo) and $+0.2$ (constrained halo): as near to zero as makes no difference.

The tallies, however, hide the most informative feature of the comparison, which is its asymmetry. When the halo wins, it tends to win big: 63 galaxies favour the free NFW halo at the strong level and 56 at the very strong level ($\dBIC<-10$), against 23 and 12 for MOND. The summed $\dBIC$ over the whole sample is $-2598$, dominated by a dozen systems with $|\dBIC|$ in the hundreds. The most extreme cases in the halo's favour are F571-8 ($\dBIC=-399$), IC~4202 ($-363$), NGC~2403 ($-355$), NGC~5907 ($-193$), UGC~05253 ($-182$), NGC~0801 ($-151$) and NGC~5055 ($-138$); in MOND's favour they are IC~2574 ($+994$), NGC~3109 ($+119$), DDO~154 ($+69$), NGC~2366 ($+38$), UGC~05986 ($+26$) and NGC~0055 ($+25$). The reader familiar with this field will recognise every name on both lists.

\begin{table}
\caption{Model comparison over the full sample (175 galaxies) and the $Q=1$ subsample (99 galaxies). The information criterion is computed as ``halo minus MOND'', so that positive values favour MOND. ``Strong'' counts galaxies with $|\Delta|>6$ (halo / MOND); ``undecided'' counts $|\Delta|\le2$.}
\label{tab:counts}
\begin{ruledtabular}
\begin{tabular}{lcccc}
Criterion & Halo & MOND & Strong & Undecided\\
 & & & halo / MOND & \\
\hline
\multicolumn{5}{l}{\textit{All 175 galaxies}}\\
BIC, free NFW & 96 & 79 & 63 / 23 & 37\\
BIC, NFW+\LCDM & 86 & 89 & 60 / 34 & 37\\
AIC, free NFW & 105 & 70 & 68 / 18 & 34\\
\multicolumn{5}{l}{\textit{$Q=1$ subsample, 99 galaxies}}\\
BIC, free NFW & 58 & 41 & 39 / 16 & \ldots\\
BIC, NFW+\LCDM & 55 & 44 & 38 / 22 & \ldots\\
AIC, free NFW & 65 & 34 & 43 / 11 & \ldots\\
\end{tabular}
\end{ruledtabular}
\end{table}

\subsection{Where each description strains}
\label{sec:where}

\begin{figure*}
\includegraphics[width=\textwidth]{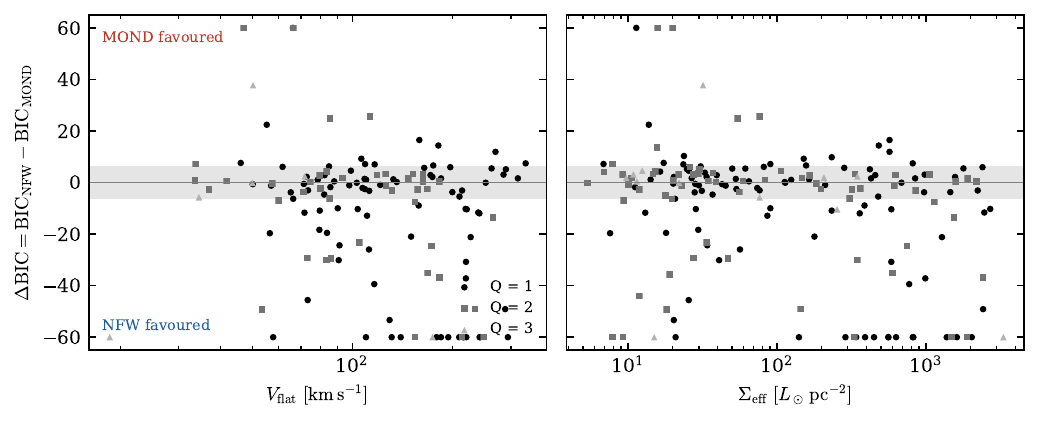}
\caption{$\dBIC$ between the free NFW halo and MOND, against flat rotation velocity (left, for the 135 galaxies with a measured $V_{\rm flat}$) and effective surface brightness (right, all galaxies), coded by quality flag. Positive values favour MOND. The grey band marks $|\dBIC|<6$, below which the evidence is not strong. Values are clipped at $\pm60$ for display; the true range extends from $-399$ to $+994$. MOND's advantage is concentrated among low-mass, low-surface-brightness galaxies, and the halo's among the massive, high-surface-brightness ones.}
\label{fig:dbic}
\end{figure*}

Figure~\ref{fig:dbic} shows $\dBIC$ against $V_{\rm flat}$ and against effective surface brightness, and Table~\ref{tab:split} splits the sample along these axes. Among the 94 galaxies with $\Sigma_{\rm eff}<100\,\Lsun\,{\rm pc}^{-2}$ MOND is preferred in 49 and the halo in 45, with median $\chin$ of $1.47$ (free halo), $1.79$ (constrained halo) and $1.80$ (MOND); among the 81 higher-surface-brightness galaxies the halo is preferred in 51 and MOND in 30. Splitting instead at $V_{\rm flat}=100\,\kms$ the halo is preferred in 32 of 54 slow rotators and 52 of 81 fast ones; but among the 40 galaxies whose rotation curves never reach a flat part---rising curves of dwarfs and low-surface-brightness disks, which is why $V_{\rm flat}$ is undefined---MOND is preferred in 28 and the halo in only 12, with a median $\dBIC=+2.6$.

The pattern is the one the two theories' critics have long described, now quantified on a uniform footing. MOND's characteristic failures are in massive, high-surface-brightness, often bulge-dominated systems---NGC~2403 (which is neither massive nor bulged, but has 73 exquisitely precise data points), NGC~5055, NGC~5907, NGC~0801, IC~4202---where the inner rotation curve rises more steeply, or the outer curve declines more, than a fixed $a_0$ acting on the observed baryons will allow; and in a handful of edge-on or distant galaxies (F571-8, UGC~05253) where the fit is driven to the edge of the inclination or distance prior. NFW's characteristic failures are in slowly rising dwarf rotation curves---IC~2574, NGC~3109, DDO~154, NGC~2366---where a cusp of any concentration $c\ge1$ produces too much velocity at small radii and too little at large ones. Nothing about the priors on $\Yd$ rescues either side: in the galaxies where MOND fails, the fit has already pushed $\Yd$ to the edge of its prior in the direction that helps, and in the galaxies where NFW fails the concentration has pinned itself at the lower bound of $c=1$, a value no cosmological halo has.

\begin{table}
\caption{Model preference by subsample, free NFW halo versus MOND. Columns give the number of galaxies, the number for which $\dBIC<0$ (halo) and $\dBIC>0$ (MOND), and the median $\chin$ for the free halo, the \LCDM-constrained halo, and MOND.}
\label{tab:split}
\begin{ruledtabular}
\begin{tabular}{lccccc}
Subsample & $n$ & Halo & MOND & \multicolumn{2}{c}{Median $\chin$}\\
 & & & & NFW / \LCDM & MOND\\
\hline
$\Sigma_{\rm eff}<100\,\Lsun\,{\rm pc}^{-2}$ & 94 & 45 & 49 & 1.47 / 1.79 & 1.80\\
$\Sigma_{\rm eff}\ge100\,\Lsun\,{\rm pc}^{-2}$ & 81 & 51 & 30 & 1.92 / 2.04 & 2.42\\
$V_{\rm flat}<100\,\kms$ & 54 & 32 & 22 & 1.28 / 1.44 & 1.68\\
$V_{\rm flat}\ge100\,\kms$ & 81 & 52 & 29 & 1.95 / 2.04 & 2.60\\
No flat part & 40 & 12 & 28 & 1.54 / 1.95 & 1.59\\
Gas-dominated\footnote{$M_{\rm HI}>0.5\,L_{[3.6]}$, i.e.\ more gas than stars for $\Yd=0.5$.} & 73 & 40 & 33 & 1.51 / 1.76 & 1.84\\
\end{tabular}
\end{ruledtabular}
\end{table}

\subsection{Halo parameters and the concentration--mass relation}
\label{sec:cM}

\begin{figure}
\includegraphics[width=\columnwidth]{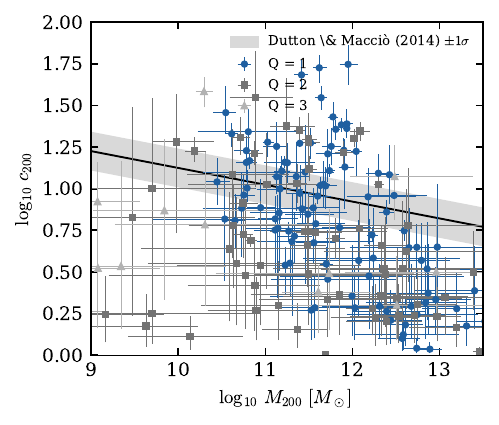}
\caption{Concentration against virial mass from the free NFW fits (posterior medians and 16--84 percentile ranges), compared with the \LCDM\ relation of \citet{DuttonMaccio2014} and its $1\sigma$ scatter (grey band). More than half of the sample lies below the band, at all masses.}
\label{fig:cM}
\end{figure}

Figure~\ref{fig:cM} plots the concentrations and virial masses from the free NFW fits over the \LCDM\ relation, Eq.~(\ref{eq:cM}). The offset $\Delta\log_{10}c$ between the fitted and predicted concentration has a median of $-0.24$ dex and a standard deviation of $0.40$~dex, to be compared with the $0.11$~dex scatter of the simulated relation. Only 30\% of the galaxies lie within $2\sigma$ ($0.22$~dex) of the relation; 53\% lie below and 17\% above. The offset does not depend on halo mass: the median is $-0.25$ dex for the 42 galaxies with $M_{200}<10^{11}\,\Msun$ and $-0.24$ dex for the 133 above it.

This is the core--cusp problem \cite{deBlok2010} in its least anecdotal form. It is usually illustrated with individual dwarfs whose inner density profiles are shallower than $r^{-1}$ \cite{Oh2011}; here it appears as a systematic demand, across the whole disk-galaxy population and at fixed virial mass, for haloes that are less centrally concentrated than \LCDM\ produces without baryonic feedback. It is also, when read together with Sec.~\ref{sec:where}, the mechanism behind the near-draw in Table~\ref{tab:counts}: the free halo's 96-to-79 advantage is bought with concentrations that the cosmological model does not supply, and when the concentration is constrained the advantage evaporates. The result agrees with what \citet{Katz2017} found with a smaller sample and a different prior on $\Yd$, and with the systematic behaviour of the NFW column in the catalogue of \citet{Li2020}; it is reassuring, and mildly disappointing, that an independent pipeline reproduces it so exactly.

\begin{figure}
\includegraphics[width=\columnwidth]{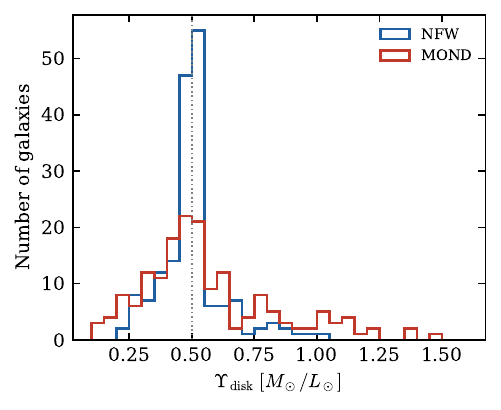}
\caption{Distribution of the posterior-median disk mass-to-light ratio in the NFW (blue) and MOND (red) fits. The prior is centred on $\Yd=0.5$ (dotted) with a width of $0.1$~dex. The halo fits are content with the prior; the MOND fits are not, because $\Yd$ is the only quantity they can move to change the shape of the rotation curve.}
\label{fig:ML}
\end{figure}

\subsection{Nuisance parameters}
\label{sec:nuisance}

The distribution of the fitted $\Yd$ (Fig.~\ref{fig:ML}) is diagnostic of how each model uses its freedom. In the NFW fits the posterior median of $\Yd$ has median $0.50$ and a 16--84\% range of $0.40$--$0.56$: the halo absorbs whatever the stars do not supply, and the mass-to-light ratio stays where the stellar-population prior put it. In the MOND fits the median is also $0.51$, but the range is $0.33$--$0.89$; the distribution is broader than the prior by a factor of three, and skewed towards high values. This is expected---in MOND the stellar mass is the only lever, and a $2\sigma$ excursion in $\Yd$ is cheap relative to a $\Delta\chi^2$ of twenty---but it is also the reason why the ``MOND has no free parameters'' claim must be read with care. The theory has none; the fit has three, and it uses them.

The same is true of distance. The dispersion of the fitted distance factor $d$ is $0.16$ in the NFW fits and $0.21$ in the MOND fits, and the number of galaxies whose fitted distance is pulled more than $2\sigma$ from the catalogued value is 10 for NFW and 28 for MOND. Almost all of the latter are galaxies with Hubble-flow distances of 20--30\% uncertainty, in which MOND asks for the galaxy to be more distant (hence more massive and, at fixed $a_0$, more Newtonian) or nearer. It is exactly this sensitivity that makes MOND a useful test: a distance-independent theory cannot be checked against the distance scale, whereas MOND's predictions for galaxies with Cepheid or tip-of-the-red-giant-branch distances are unforgiving. Among the 21 galaxies with $Q=1$ and a Cepheid or tip-of-the-red-giant-branch distance, where the distance uncertainty is 5--10\%, the median $\chin$ of MOND is $2.1$ against $1.5$ for the free halo, and four of them---NGC~0891, NGC~3198, NGC~5907 and NGC~7814, all massive, high-surface-brightness spirals---still have their distance pulled by 2--3$\sigma$. For a distance known to 10\% that is not an artefact of a wide prior; it is a measurement of how much these galaxies would have to move to make a fixed $a_0$ work.

\subsection{The radial acceleration relation}
\label{sec:rar}

\begin{figure}
\includegraphics[width=\columnwidth]{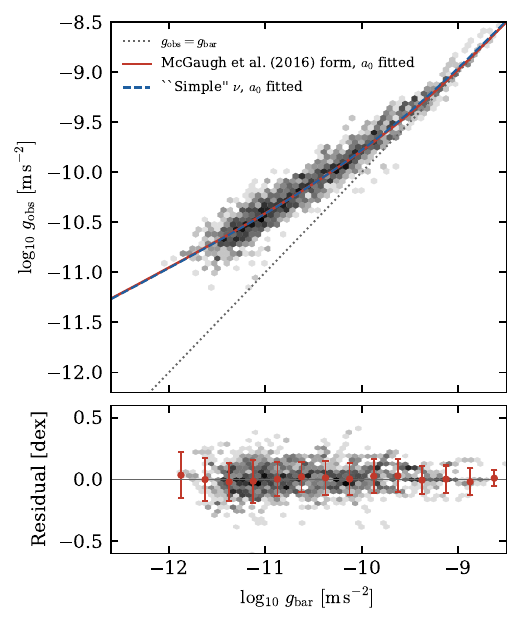}
\caption{The radial acceleration relation for 2803 points from all 175 galaxies, with $\Yd=0.5$, $\Yb=0.7$ and the published distances and inclinations. Top: $g_{\rm obs}$ against $g_{\rm bar}$ (grey density), with the one-parameter fits of Eq.~(\ref{eq:mond}) using the interpolation function of \citet{McGaugh2016} (red) and the simple function (blue, dashed); the dotted line is $g_{\rm obs}=g_{\rm bar}$. Bottom: residuals from the red curve, with binned medians and standard deviations. The relation is tight at $0.14$~dex, and the residuals show no trend with $g_{\rm bar}$ at the $0.03$~dex level.}
\label{fig:rar}
\end{figure}

The RAR \cite{McGaugh2016,Lelli2017} is reconstructed from the full sample by computing $g_{\rm obs}=V_{\rm obs}^2/R$ and $g_{\rm bar}=V_{\rm bar}^2/R$ at every rotation-curve point with the fiducial $\Yd=0.5$, $\Yb=0.7$ and the catalogued distances and inclinations, and retaining points with $\delta V_{\rm obs}/V_{\rm obs}<0.1$ and $g_{\rm bar}>0$, the same cuts as the discovery paper. This leaves 2803 points. A single-parameter fit of Eq.~(\ref{eq:mond}) in the logarithm gives
\begin{equation}
a_0=(1.13\pm0.02)\times10^{-10}\,\mss
\label{eq:a0}
\end{equation}
with the interpolation function of \citet{McGaugh2016} and $(1.10\pm0.02)\times10^{-10}\,\mss$ with the simple function, the quoted errors being statistical only. The observed scatter about the relation is $0.144$~dex for the full sample and $0.128$~dex for the $Q=1$ galaxies, and the binned residuals (Fig.~\ref{fig:rar}, lower panel) are flat to within $0.03$~dex over three decades in $g_{\rm bar}$. These numbers are consistent with $a_0=(1.20\pm0.02)\times10^{-10}$ and a scatter of $0.13$~dex reported by \citet{McGaugh2016}, the slight offset in $a_0$ being within what one expects from the different handling of the $Q=3$ galaxies and of the inclination cut; more importantly, they show that the relation survives the inclusion of every galaxy the database contains.

The RAR is a statement about the population. The scatter of $0.14$~dex is comparable to what the observational errors on $V_{\rm obs}$, distance, inclination and $\Yd$ produce when propagated \cite{Lelli2017,Desmond2023}, so the relation is, as far as this data set can tell, consistent with having no intrinsic scatter at all. That is the empirical fact any theory of galactic dynamics has to reproduce. In MOND it is the theory. In \LCDM\ it must be an outcome of galaxy formation, and the fact that the free NFW fits reproduce it only by adjusting $c$ galaxy by galaxy, in a direction that the concentration--mass relation does not sanction, is the sharpest form of the challenge.

\begin{figure*}
\includegraphics[width=\textwidth]{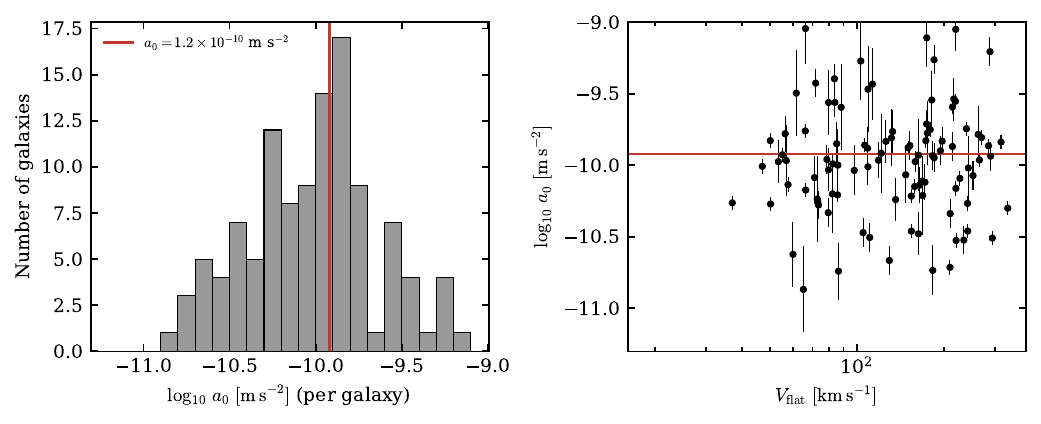}
\caption{The acceleration constant preferred by individual galaxies when $a_0$ is allowed to float in the MOND fit, together with $\Yd$, $\Yb$, distance and inclination under their usual priors. Left: distribution of $\log_{10}a_0$ for the 121 galaxies in which the profile-likelihood uncertainty on $\log_{10}a_0$ is below $0.3$~dex; the red line is the canonical $1.2\times10^{-10}\,\mss$. Right: the same against $V_{\rm flat}$. The scatter is far larger than the uncertainties and, unlike the RAR residuals, is not consistent with a single value.}
\label{fig:a0}
\end{figure*}

\subsection{Is there one acceleration scale?}
\label{sec:a0}

\citet{Rodrigues2018} argued that if $a_0$ is treated as a free parameter in each galaxy, the SPARC galaxies do not agree on its value, and that this rules out MOND as a fundamental theory at high confidence. \citet{McGaugh2018} and \citet{Kroupa2018} replied that this procedure conflates $a_0$ with the nuisance parameters, principally $\Yd$ and distance, with which it is degenerate, and that the tightness of the RAR is the relevant fact. Both sides are, in a sense, right, and the point of this section is to show both results in the same pipeline.

Refitting each galaxy with $\log_{10}a_0$ as an additional free parameter (uniform prior between $10^{-12}$ and $10^{-8}\,\mss$) and the nuisance parameters as before, and estimating the uncertainty on $\log_{10}a_0$ by profiling the posterior, yields the distribution of Fig.~\ref{fig:a0}. For the 121 galaxies whose profile constrains $\log_{10}a_0$ to better than $0.3$~dex, the median is $1.02\times10^{-10}\,\mss$, gratifyingly close to the canonical value, but the median absolute deviation is $0.36$~dex and the standard deviation $0.58$~dex. A fit of a constant to these 121 values with their profile errors gives $\chi^2=3120$ for 120 degrees of freedom, and the implied intrinsic scatter is $0.55$~dex. The per-galaxy $a_0$ is uncorrelated with luminosity or $V_{\rm flat}$ (Spearman $\rho\simeq0.1$, $p>0.3$) but correlates with effective surface brightness ($\rho=0.33$, $p=3\times10^{-4}$), in the sense that higher-surface-brightness galaxies prefer a larger $a_0$.

Taken naively, this is the result of \citet{Rodrigues2018}. Taken together with Sec.~\ref{sec:rar}, it says something more interesting. The same data, with the same $\Yd$ prior, produce a population relation whose scatter is explained by the errors and a per-galaxy constant whose scatter is not. The difference lies in what the two procedures allow to vary. In the RAR, $\Yd$ and the distance are held at their fiducial values and $a_0$ is the only free quantity, so the errors on $\Yd$ and $D$ enter as scatter about a fixed curve. In the per-galaxy fits, $\Yd$ and $D$ are free within their priors, and because $a_0$, $\Yd$ and $D$ enter the low-acceleration regime in nearly degenerate combinations---a galaxy with more stellar mass or at a greater distance needs a smaller $a_0$ to produce the same rotation curve---the fit can trade one for the other at little cost in $\chi^2$. What looks like a spread in $a_0$ is, to a large extent, a spread in the combination the data actually constrain, projected onto one axis; the correlation with surface brightness, the quantity most directly tied to $\Yd$, is the fingerprint of that projection. This does not prove that $a_0$ is universal. It shows that the per-galaxy test as usually posed is not sensitive enough to tell, and that the only way to make it sensitive is to pin the nuisance parameters independently: with resolved stellar-population mass-to-light ratios and with distances of a few per cent, both of which are becoming available for the nearest galaxies.

\section{Discussion}
\label{sec:discussion}

\paragraph*{What the draw means.}
The headline count---96 to 79 for the free halo, 86 to 89 with the cosmological prior---will be read by partisans of either side as support. It should not be. The free NFW halo is a two-parameter function fitted to curves with a median of 14 points, and it is preferred over a zero-parameter prediction in barely more than half the sample; the cosmologically constrained halo is not preferred at all. Conversely, MOND is preferred over a halo with two extra parameters in nearly half the sample, which is remarkable for a theory with nothing to adjust beyond the nuisance parameters everyone shares, but it also fails outright, at $|\dBIC|$ of hundreds, in a dozen galaxies whose data are among the best in the database. A theory that is exactly right would not do that.

The more useful reading is that the two descriptions are good at different galaxies, and that the boundary between their domains is physically meaningful. NFW is good where the baryons dominate the inner dynamics and the halo is needed only to hold up the outer curve; MOND is good where the baryons are diffuse and the whole curve lies in the low-acceleration regime. Any theory that hopes to supersede both must reproduce the halo's behaviour in the first regime and MOND's in the second, which is a more specific requirement than either paradigm's advocates usually state.

\paragraph*{The concentration offset.}
The most robust result of this work is the systematic $0.24$~dex deficit in NFW concentration relative to \LCDM, present at all halo masses. In the standard picture this deficit is attributed to the redistribution of dark matter by supernova-driven gas outflows, which flatten cusps in haloes where the stellar-to-halo mass ratio is a few per mille to a few per cent \cite{DiCintio2014}, and to the adiabatic response of the halo to the disk, which in principle acts the other way. Feedback-modified profiles do fit SPARC better \cite{Katz2017,Li2020}. But the deficit found here is as large for haloes of $10^{12}\,\Msun$, where the feedback models predict cusps to survive, as for haloes of $10^{10}\,\Msun$, where they predict cores. Either the feedback prescriptions need to act over a wider mass range than the simulations that motivate them, or something else is going on. This deserves a dedicated study with the mass-dependent profiles, which the present pipeline can accommodate trivially.

\paragraph*{The two faces of $a_0$.}
The reconciliation offered in Sec.~\ref{sec:a0} is, admittedly, the kind of argument that both camps will find partially satisfactory. To a proponent of MOND it says that the per-galaxy spread in $a_0$ is an artefact of nuisance-parameter degeneracies, which is what they have argued. To a sceptic it says that the tight RAR is likewise a product of holding those nuisance parameters fixed, and that a universal $a_0$ has not been demonstrated at the level of individual systems, which is what they have argued. The resolution is not rhetorical but observational: $\Yd$ and $D$ must be measured, not fitted. Where the distance at least has been measured to a few per cent, the outcome in this sample is already suggestive: the low-mass galaxies with Cepheid or TRGB distances are fitted by MOND without incident, whereas four massive spirals (Sec.~\ref{sec:nuisance}) can only be fitted by moving them 2--3$\sigma$ along the line of sight. If independent mass-to-light ratios for those four confirm the fiducial $\Yd$, the tension is with $a_0$ itself; if they do not, it is with the stellar-population models. Either answer is more informative than another hundred rotation curves with Hubble-flow distances.

\paragraph*{Limitations.}
Several simplifications limit the reach of these results. The gas velocities are taken as published, without the $\sim$10\% uncertainty in the \HI-to-total-gas conversion. The MOND model is algebraic and ignores the external field effect, which for the isolated galaxies in SPARC is a small correction but not zero. The NFW profile is used without the adiabatic contraction that a real disk would induce, which would make the cusp problem worse, not better. The BIC is a crude proxy for the evidence. And, most seriously, the $\Yd$ prior of $0.1$~dex is a choice; broadening it to $0.2$~dex improves the MOND fits considerably, as \citet{Li2018} found, but at the price of mass-to-light ratios that stellar-population models do not support. Every one of these choices is stated, and the code and full results are available on request, so that anyone who prefers a different choice can see what it costs.

\section{Conclusions}
\label{sec:conclusions}

The 175 rotation curves of the SPARC database have been fitted, under identical assumptions, with an NFW dark-matter halo, an NFW halo obeying the \LCDM\ concentration--mass relation, and MOND with a fixed acceleration constant. The comparison yields the following.

\begin{enumerate}
\item By the Bayesian information criterion, the free NFW halo is preferred in 96 galaxies and MOND in 79; with the cosmological concentration prior the count is 86 to 89. Neither description wins. The halo's advantage is concentrated in massive, high-surface-brightness systems and is purchased with concentrations that \LCDM\ does not predict; MOND's advantage is concentrated in slowly rising dwarf rotation curves, where cusps of any concentration fail.
\item The fitted NFW concentrations lie a median $0.24$~dex below the Planck-cosmology relation, with 53\% of galaxies more than $2\sigma$ low, at all halo masses from $10^{10}$ to $10^{13}\,\Msun$. This is the core--cusp problem as a population statistic.
\item MOND fits use their nuisance parameters harder than halo fits do: the fitted disk mass-to-light ratio spreads over $0.33$--$0.89$ against $0.40$--$0.56$ for NFW, and 28 galaxies have their distance moved by more than $2\sigma$ against 10 for NFW.
\item The radial acceleration relation from the full sample has $a_0=(1.13\pm0.02)\times10^{-10}\,\mss$ and a scatter of $0.14$~dex ($0.13$~dex for $Q=1$), consistent with no intrinsic scatter.
\item The acceleration constant fitted galaxy by galaxy has a median of $1.0\times10^{-10}\,\mss$ but an intrinsic spread of $0.55$~dex that correlates with surface brightness. This is a consequence of the degeneracy between $a_0$, $\Yd$ and distance, not evidence for or against a universal constant; resolving it requires independent measurements of the nuisance parameters, not more rotation curves.
\end{enumerate}

The picture that emerges is of two descriptions, each of which is right about something the other struggles to explain. The dark-matter halo explains why the dynamics of galaxies is decoupled from their baryons where the baryons are dense, and fails to explain, without help, why the haloes of real galaxies are so much less concentrated than simulated ones. MOND explains, with one constant, why the dynamics of galaxies is so tightly coupled to their baryons where the baryons are diffuse, and fails to explain why that coupling breaks in the systems with the best data. Rotation curves alone will not decide which of these is the deeper failure. They have, however, made the two failures precise enough to be worth taking seriously, which after fifty years is progress.

\begin{acknowledgments}
This work made use of the SPARC database (Spitzer Photometry and Accurate Rotation Curves), compiled and maintained by F.~Lelli, S.~S.~McGaugh and J.~M.~Schombert, and of the original \HI\ and H$\alpha$ observations collected therein, which represent decades of effort by many observers. The analysis used the \textsc{numpy}, \textsc{scipy}, \textsc{matplotlib} and \textsc{emcee} packages.
\end{acknowledgments}

\appendix

\section{Fits for all galaxies}
\label{app:table}

Table~\ref{tab:all1} lists, for every galaxy, the quality flag, the number of rotation-curve points, the reduced $\chi^2$ of the three models at their maximum-a-posteriori points, the posterior-median concentration and virial mass from the free NFW fit, the posterior-median disk mass-to-light ratio from the MOND fit, and $\dBIC$ between the free NFW halo and MOND (positive favours MOND). Galaxies without a bulge have $\Yb$ pinned by its prior and it is not counted as a free parameter. All quantities are as defined in the text.

\begin{table*}
\caption{Fits for all 175 SPARC galaxies (part 1 of 4). See Appendix~\ref{app:table} for column definitions.}
\label{tab:all1}
\begin{ruledtabular}
\begin{tabular}{lccccccccc}
Galaxy & $Q$ & $N$ & $\chin({\rm NFW})$ & $\log_{10}c$ & $\log_{10}M_{200}$ & $\chin(\Lambda{\rm CDM})$ & $\chin({\rm MOND})$ & $\Yd({\rm MOND})$ & $\dBIC$\\
 & & & & & $[\Msun]$ & & & $[\Msun/\Lsun]$ & \\
\hline
CamB & 2 & 9 & 8.64 & 0.18 & 9.63 & 11.79 & 20.62 & 0.12 & -84.8\\
D512-2 & 2 & 4 & 0.62 & 0.54 & 10.95 & 1.23 & 0.35 & 0.48 & +3.0\\
D564-8 & 2 & 6 & 3.73 & 0.27 & 10.89 & 11.42 & 2.72 & 0.38 & -0.9\\
D631-7 & 1 & 16 & 6.00 & 0.10 & 12.57 & 6.53 & 12.85 & 0.20 & -95.6\\
DDO064 & 1 & 14 & 0.67 & 0.36 & 12.00 & 1.09 & 0.34 & 0.49 & +7.6\\
DDO154 & 2 & 12 & 13.64 & 0.69 & 10.83 & 13.90 & 3.45 & 0.20 & +69.4\\
DDO161 & 1 & 31 & 1.02 & 0.27 & 11.53 & 1.24 & 1.42 & 0.24 & -6.3\\
DDO168 & 2 & 10 & 14.27 & 0.17 & 13.20 & 14.68 & 17.90 & 0.45 & -49.4\\
DDO170 & 2 & 8 & 4.32 & 0.72 & 10.75 & 4.52 & 4.82 & 0.78 & -7.0\\
ESO079-G014 & 1 & 15 & 4.67 & 0.33 & 12.96 & 4.95 & 3.79 & 0.54 & +6.6\\
ESO116-G012 & 1 & 15 & 3.02 & 0.77 & 11.85 & 3.03 & 2.37 & 0.36 & +7.2\\
ESO444-G084 & 2 & 7 & 1.49 & 0.74 & 11.49 & 1.75 & 3.33 & 0.42 & -6.4\\
ESO563-G021 & 1 & 30 & 19.05 & 0.13 & 13.85 & 19.38 & 17.83 & 0.61 & +1.6\\
F561-1 & 3 & 6 & 1.20 & 0.87 & 9.84 & 1.21 & 1.63 & 0.52 & -0.1\\
F563-1 & 1 & 17 & 1.34 & 0.85 & 11.49 & 1.34 & 1.47 & 0.66 & +1.1\\
F563-V1 & 3 & 6 & 0.99 & 0.52 & 9.08 & 1.61 & 0.91 & 0.46 & +1.8\\
F563-V2 & 1 & 10 & 1.90 & 0.58 & 12.24 & 1.98 & 1.01 & 0.60 & +7.1\\
F565-V2 & 2 & 7 & 0.94 & 0.34 & 12.51 & 1.98 & 0.42 & 0.51 & +4.1\\
F567-2 & 3 & 5 & 1.78 & 0.78 & 10.31 & 1.76 & 2.27 & 0.55 & +0.5\\
F568-1 & 1 & 12 & 1.19 & 0.65 & 12.65 & 1.20 & 1.24 & 0.65 & +2.2\\
F568-3 & 1 & 18 & 3.88 & 0.12 & 12.58 & 4.34 & 3.07 & 0.46 & +10.3\\
F568-V1 & 1 & 15 & 0.28 & 1.16 & 11.25 & 0.31 & 0.95 & 0.78 & -3.2\\
F571-8 & 1 & 13 & 4.31 & 0.96 & 12.48 & 4.48 & 43.82 & 0.23 & -398.5\\
F571-V1 & 2 & 7 & 0.78 & 0.49 & 11.50 & 1.04 & 0.26 & 0.49 & +4.4\\
F574-1 & 1 & 14 & 2.18 & 0.88 & 11.42 & 2.19 & 2.36 & 0.72 & -1.1\\
F574-2 & 3 & 5 & 0.13 & 0.53 & 9.34 & 0.40 & 0.08 & 0.48 & +3.2\\
F579-V1 & 1 & 14 & 0.25 & 1.34 & 10.80 & 0.34 & 3.05 & 0.66 & -26.0\\
F583-1 & 1 & 25 & 1.78 & 0.68 & 11.56 & 1.75 & 1.99 & 0.98 & -1.8\\
F583-4 & 1 & 12 & 0.21 & 0.46 & 11.72 & 0.35 & 0.14 & 0.49 & +5.2\\
IC2574 & 2 & 34 & 37.14 & 0.00 & 11.69 & 37.19 & 2.91 & 0.10 & +993.8\\
IC4202 & 1 & 32 & 20.54 & 1.36 & 11.94 & 20.53 & 32.28 & 0.95 & -362.7\\
KK98-251 & 2 & 15 & 1.51 & 0.11 & 10.13 & 4.04 & 1.11 & 0.44 & +7.3\\
NGC0024 & 1 & 29 & 0.92 & 1.28 & 11.03 & 0.80 & 0.76 & 1.02 & +9.2\\
NGC0055 & 2 & 21 & 2.84 & 0.28 & 12.23 & 3.00 & 1.49 & 0.18 & +24.8\\
NGC0100 & 1 & 21 & 0.89 & 0.29 & 12.68 & 1.04 & 1.11 & 0.29 & +0.5\\
NGC0247 & 2 & 26 & 2.02 & 0.74 & 11.46 & 2.07 & 3.14 & 0.76 & -23.3\\
NGC0289 & 2 & 28 & 1.95 & 0.70 & 11.81 & 2.04 & 2.05 & 0.90 & +0.4\\
NGC0300 & 2 & 25 & 0.74 & 0.74 & 11.58 & 0.74 & 0.89 & 0.40 & +1.8\\
NGC0801 & 1 & 13 & 6.77 & 0.18 & 12.61 & 7.78 & 21.06 & 0.57 & -151.3\\
NGC0891 & 1 & 18 & 3.31 & 1.36 & 11.81 & 4.16 & 7.86 & 0.33 & -64.6\\
NGC1003 & 1 & 36 & 2.48 & 0.55 & 11.70 & 2.48 & 4.82 & 0.37 & -75.2\\
NGC1090 & 1 & 24 & 2.72 & 1.03 & 11.67 & 2.75 & 2.50 & 0.75 & +5.7\\
NGC1705 & 3 & 14 & 0.15 & 1.58 & 10.30 & 0.44 & 0.39 & 1.23 & +2.4\\
NGC2366 & 3 & 26 & 3.50 & 0.38 & 11.61 & 3.66 & 1.83 & 0.26 & +37.8\\
NGC2403 & 1 & 73 & 8.97 & 1.27 & 11.40 & 9.02 & 13.91 & 0.50 & -355.3\\
\end{tabular}
\end{ruledtabular}
\end{table*}

\begin{table*}
\caption{Fits for all 175 SPARC galaxies (part 2 of 4). See Appendix~\ref{app:table} for column definitions.}
\label{tab:all2}
\begin{ruledtabular}
\begin{tabular}{lccccccccc}
Galaxy & $Q$ & $N$ & $\chin({\rm NFW})$ & $\log_{10}c$ & $\log_{10}M_{200}$ & $\chin(\Lambda{\rm CDM})$ & $\chin({\rm MOND})$ & $\Yd({\rm MOND})$ & $\dBIC$\\
 & & & & & $[\Msun]$ & & & $[\Msun/\Lsun]$ & \\
\hline
NGC2683 & 2 & 11 & 1.50 & 1.30 & 11.50 & 1.79 & 2.83 & 0.53 & -7.5\\
NGC2841 & 1 & 50 & 1.48 & 0.75 & 12.59 & 1.48 & 1.52 & 0.76 & +3.1\\
NGC2903 & 1 & 34 & 6.34 & 1.68 & 11.42 & 6.52 & 8.84 & 0.26 & -83.1\\
NGC2915 & 2 & 30 & 0.97 & 1.21 & 10.87 & 1.01 & 2.26 & 0.81 & -30.1\\
NGC2955 & 1 & 24 & 4.02 & 1.09 & 12.30 & 4.19 & 4.13 & 0.35 & -3.8\\
NGC2976 & 2 & 27 & 0.84 & 0.25 & 9.70 & 1.20 & 1.30 & 0.35 & -6.2\\
NGC2998 & 1 & 13 & 3.05 & 1.13 & 11.91 & 3.40 & 3.50 & 0.83 & -5.4\\
NGC3109 & 1 & 25 & 9.91 & 0.04 & 12.74 & 10.57 & 3.88 & 0.23 & +119.4\\
NGC3198 & 1 & 43 & 1.46 & 1.02 & 11.63 & 1.46 & 2.10 & 0.74 & -21.0\\
NGC3521 & 1 & 41 & 0.25 & 0.39 & 13.40 & 0.29 & 0.51 & 0.42 & -3.1\\
NGC3726 & 2 & 12 & 2.73 & 0.24 & 12.71 & 3.34 & 2.95 & 0.46 & -2.5\\
NGC3741 & 1 & 21 & 0.41 & 0.54 & 11.23 & 0.52 & 0.74 & 0.32 & -0.6\\
NGC3769 & 2 & 12 & 0.94 & 1.03 & 11.36 & 0.98 & 0.96 & 0.41 & +2.9\\
NGC3877 & 2 & 13 & 5.91 & 1.28 & 11.50 & 6.34 & 8.75 & 0.45 & -35.1\\
NGC3893 & 1 & 10 & 1.01 & 1.21 & 11.72 & 1.50 & 1.07 & 0.43 & +2.2\\
NGC3917 & 1 & 17 & 3.34 & 0.26 & 12.39 & 4.11 & 3.25 & 0.60 & +0.2\\
NGC3949 & 2 & 7 & 0.75 & 0.62 & 12.62 & 0.74 & 0.58 & 0.43 & +3.1\\
NGC3953 & 1 & 8 & 0.51 & 1.10 & 11.46 & 0.89 & 3.21 & 0.57 & -10.4\\
NGC3972 & 1 & 10 & 1.80 & 0.52 & 12.86 & 1.81 & 1.76 & 0.50 & +1.2\\
NGC3992 & 1 & 9 & 0.72 & 1.22 & 12.04 & 1.08 & 3.21 & 0.75 & -12.0\\
NGC4010 & 2 & 12 & 3.09 & 0.35 & 13.06 & 3.21 & 2.60 & 0.37 & +3.2\\
NGC4013 & 2 & 36 & 0.80 & 0.23 & 12.97 & 0.97 & 1.74 & 0.32 & -24.5\\
NGC4051 & 2 & 7 & 1.72 & 1.02 & 11.02 & 2.04 & 2.52 & 0.44 & -2.7\\
NGC4068 & 2 & 6 & 7.17 & 0.30 & 11.15 & 13.20 & 1.89 & 0.40 & +5.1\\
NGC4085 & 2 & 7 & 9.80 & 0.50 & 13.39 & 9.87 & 6.64 & 0.36 & -3.1\\
NGC4088 & 1 & 12 & 0.58 & 0.32 & 12.84 & 0.68 & 0.68 & 0.38 & +2.9\\
NGC4100 & 1 & 24 & 0.99 & 1.27 & 11.52 & 1.08 & 1.63 & 0.71 & -8.9\\
NGC4138 & 2 & 7 & 4.49 & 1.38 & 11.24 & 5.98 & 2.29 & 0.53 & +1.5\\
NGC4157 & 1 & 17 & 0.49 & 0.37 & 12.87 & 0.59 & 0.72 & 0.42 & +1.6\\
NGC4183 & 1 & 23 & 0.20 & 1.07 & 11.13 & 0.20 & 1.13 & 0.82 & -12.9\\
NGC4214 & 2 & 14 & 0.66 & 1.55 & 10.89 & 0.97 & 1.23 & 0.46 & -2.3\\
NGC4217 & 1 & 19 & 3.68 & 0.95 & 12.18 & 3.55 & 2.62 & 1.01 & +14.4\\
NGC4389 & 3 & 6 & 21.89 & 0.30 & 12.51 & 21.84 & 7.85 & 0.30 & +1.9\\
NGC4559 & 1 & 32 & 0.24 & 1.07 & 11.35 & 0.27 & 0.50 & 0.51 & -0.9\\
NGC5005 & 1 & 18 & 0.10 & 0.65 & 12.97 & 0.10 & 0.11 & 0.51 & +5.5\\
NGC5033 & 1 & 22 & 3.51 & 1.43 & 11.78 & 3.44 & 7.89 & 0.96 & -79.7\\
NGC5055 & 1 & 28 & 2.58 & 1.11 & 11.74 & 2.53 & 8.17 & 0.51 & -138.1\\
NGC5371 & 1 & 19 & 5.06 & 1.55 & 11.64 & 5.17 & 12.29 & 2.43 & -119.9\\
NGC5585 & 1 & 24 & 6.65 & 0.97 & 11.40 & 6.66 & 6.79 & 0.23 & -10.0\\
NGC5907 & 1 & 19 & 5.24 & 1.38 & 11.87 & 5.73 & 17.01 & 0.61 & -193.0\\
NGC5985 & 1 & 33 & 1.92 & 1.75 & 11.95 & 2.49 & 7.10 & 0.55 & -147.0\\
NGC6015 & 2 & 44 & 8.76 & 1.35 & 11.40 & 8.87 & 10.75 & 1.06 & -91.5\\
NGC6195 & 1 & 23 & 2.08 & 0.28 & 13.20 & 2.10 & 2.20 & 0.31 & -0.1\\
NGC6503 & 1 & 31 & 1.55 & 1.16 & 11.22 & 1.56 & 3.09 & 0.44 & -39.4\\
NGC6674 & 1 & 15 & 2.90 & 0.04 & 12.89 & 4.17 & 10.11 & 0.89 & -79.6\\
\end{tabular}
\end{ruledtabular}
\end{table*}

\begin{table*}
\caption{Fits for all 175 SPARC galaxies (part 3 of 4). See Appendix~\ref{app:table} for column definitions.}
\label{tab:all3}
\begin{ruledtabular}
\begin{tabular}{lccccccccc}
Galaxy & $Q$ & $N$ & $\chin({\rm NFW})$ & $\log_{10}c$ & $\log_{10}M_{200}$ & $\chin(\Lambda{\rm CDM})$ & $\chin({\rm MOND})$ & $\Yd({\rm MOND})$ & $\dBIC$\\
 & & & & & $[\Msun]$ & & & $[\Msun/\Lsun]$ & \\
\hline
NGC6789 & 2 & 4 & 4.07 & 0.78 & 12.64 & 4.23 & 6.18 & 0.63 & +0.7\\
NGC6946 & 1 & 58 & 1.68 & 1.02 & 11.69 & 1.68 & 1.46 & 0.58 & +16.5\\
NGC7331 & 1 & 36 & 0.81 & 0.86 & 12.39 & 0.80 & 1.34 & 0.29 & -11.6\\
NGC7793 & 1 & 46 & 0.90 & 0.68 & 11.31 & 0.92 & 0.81 & 0.55 & +9.8\\
NGC7814 & 1 & 18 & 0.55 & 1.23 & 11.90 & 0.64 & 1.62 & 1.18 & -10.2\\
PGC51017 & 3 & 6 & 7.80 & 0.27 & 8.78 & 7.55 & 52.73 & 0.15 & -146.8\\
UGC00128 & 1 & 22 & 3.57 & 0.96 & 11.60 & 3.57 & 6.33 & 2.47 & -53.3\\
UGC00191 & 1 & 9 & 5.94 & 0.89 & 10.95 & 6.06 & 3.79 & 1.14 & +5.4\\
UGC00634 & 2 & 4 & 5.53 & 0.66 & 11.48 & 6.90 & 2.33 & 0.48 & +6.0\\
UGC00731 & 1 & 12 & 0.46 & 1.00 & 10.79 & 0.44 & 5.97 & 2.60 & -45.6\\
UGC00891 & 2 & 5 & 9.07 & 0.20 & 12.39 & 10.96 & 24.04 & 0.34 & -35.8\\
UGC01230 & 1 & 11 & 1.62 & 0.82 & 10.53 & 1.62 & 3.10 & 0.78 & -10.3\\
UGC01281 & 1 & 25 & 1.08 & 0.21 & 12.44 & 1.46 & 0.25 & 0.41 & +22.4\\
UGC02023 & 2 & 5 & 2.14 & 0.48 & 10.78 & 2.42 & 1.11 & 0.49 & +3.1\\
UGC02259 & 2 & 8 & 1.40 & 1.31 & 10.71 & 1.60 & 7.55 & 1.21 & -29.4\\
UGC02455 & 3 & 8 & 6.63 & 0.50 & 12.96 & 6.82 & 6.89 & 0.30 & -10.4\\
UGC02487 & 1 & 17 & 5.24 & 1.08 & 12.43 & 5.12 & 4.30 & 1.66 & +7.4\\
UGC02885 & 1 & 19 & 0.99 & 0.65 & 12.74 & 1.00 & 0.91 & 0.45 & +5.1\\
UGC02916 & 2 & 43 & 11.37 & 1.22 & 11.90 & 11.38 & 11.92 & 1.36 & -36.8\\
UGC02953 & 2 & 115 & 5.64 & 1.02 & 12.30 & 5.63 & 5.74 & 0.53 & -13.6\\
UGC03205 & 1 & 48 & 3.07 & 1.73 & 11.62 & 3.59 & 3.80 & 0.64 & -30.8\\
UGC03546 & 1 & 30 & 0.95 & 1.25 & 11.76 & 0.98 & 0.91 & 0.64 & +6.0\\
UGC03580 & 2 & 47 & 2.33 & 1.12 & 11.50 & 2.33 & 2.42 & 0.28 & -1.1\\
UGC04278 & 1 & 25 & 1.35 & 0.18 & 13.31 & 1.51 & 2.63 & 0.56 & -24.4\\
UGC04305 & 3 & 22 & 1.80 & 0.92 & 9.07 & 1.85 & 2.24 & 0.83 & -5.8\\
UGC04325 & 1 & 8 & 5.33 & 1.23 & 10.78 & 5.09 & 10.05 & 1.00 & -30.1\\
UGC04483 & 2 & 8 & 1.18 & 0.64 & 10.59 & 1.39 & 0.86 & 0.43 & +3.4\\
UGC04499 & 1 & 9 & 1.79 & 0.75 & 11.11 & 1.78 & 1.56 & 0.53 & +2.2\\
UGC05005 & 1 & 11 & 0.33 & 0.28 & 12.03 & 0.44 & 0.27 & 0.46 & +4.6\\
UGC05253 & 2 & 73 & 2.41 & 1.35 & 12.09 & 2.48 & 5.11 & 0.53 & -182.0\\
UGC05414 & 1 & 6 & 3.95 & 0.29 & 12.31 & 5.83 & 1.24 & 0.41 & +3.8\\
UGC05716 & 2 & 12 & 2.48 & 0.91 & 10.74 & 2.55 & 5.74 & 1.39 & -29.3\\
UGC05721 & 1 & 23 & 1.09 & 1.46 & 10.55 & 1.19 & 1.84 & 0.64 & -10.9\\
UGC05750 & 1 & 11 & 1.45 & 0.29 & 11.57 & 2.12 & 0.79 & 0.49 & +7.2\\
UGC05764 & 2 & 10 & 10.57 & 1.23 & 10.19 & 10.47 & 17.34 & 3.02 & -63.9\\
UGC05829 & 2 & 11 & 0.13 & 0.36 & 11.85 & 0.49 & 0.48 & 0.63 & +1.7\\
UGC05918 & 2 & 8 & 0.29 & 0.78 & 10.63 & 0.40 & 1.03 & 0.53 & -0.1\\
UGC05986 & 2 & 15 & 5.86 & 0.30 & 12.75 & 6.65 & 3.20 & 0.33 & +25.7\\
UGC05999 & 2 & 5 & 6.56 & 0.52 & 12.36 & 6.11 & 5.41 & 0.49 & -1.0\\
UGC06399 & 1 & 9 & 1.14 & 0.57 & 12.07 & 1.33 & 0.45 & 0.53 & +6.3\\
UGC06446 & 1 & 17 & 0.27 & 1.16 & 10.78 & 0.28 & 0.97 & 1.03 & -4.7\\
UGC06614 & 1 & 13 & 0.27 & 0.57 & 12.79 & 0.29 & 1.19 & 0.50 & -3.7\\
UGC06628 & 2 & 7 & 0.26 & 1.00 & 9.70 & 0.37 & 0.94 & 0.51 & +0.6\\
UGC06667 & 1 & 9 & 2.28 & 0.79 & 11.58 & 2.36 & 5.51 & 1.11 & -19.5\\
UGC06786 & 1 & 45 & 0.69 & 1.39 & 11.94 & 0.82 & 1.75 & 0.32 & -37.2\\
\end{tabular}
\end{ruledtabular}
\end{table*}

\begin{table*}
\caption{Fits for all 175 SPARC galaxies (part 4 of 4). See Appendix~\ref{app:table} for column definitions.}
\label{tab:all4}
\begin{ruledtabular}
\begin{tabular}{lccccccccc}
Galaxy & $Q$ & $N$ & $\chin({\rm NFW})$ & $\log_{10}c$ & $\log_{10}M_{200}$ & $\chin(\Lambda{\rm CDM})$ & $\chin({\rm MOND})$ & $\Yd({\rm MOND})$ & $\dBIC$\\
 & & & & & $[\Msun]$ & & & $[\Msun/\Lsun]$ & \\
\hline
UGC06787 & 2 & 71 & 18.35 & 0.02 & 13.46 & 18.59 & 20.15 & 0.46 & -148.7\\
UGC06818 & 2 & 8 & 5.84 & 0.24 & 12.53 & 7.18 & 5.09 & 0.31 & -3.7\\
UGC06917 & 1 & 11 & 1.16 & 0.88 & 11.55 & 1.15 & 1.30 & 0.54 & +1.4\\
UGC06923 & 2 & 6 & 3.67 & 0.48 & 12.38 & 4.17 & 1.49 & 0.42 & +2.8\\
UGC06930 & 1 & 10 & 0.44 & 1.00 & 11.17 & 0.44 & 1.27 & 0.63 & -2.1\\
UGC06973 & 3 & 9 & 2.30 & 1.07 & 12.49 & 3.30 & 16.57 & 0.16 & -71.5\\
UGC06983 & 1 & 17 & 0.74 & 1.11 & 11.19 & 0.76 & 1.22 & 0.78 & -2.5\\
UGC07089 & 2 & 12 & 0.36 & 0.23 & 12.26 & 0.97 & 0.41 & 0.36 & +3.8\\
UGC07125 & 1 & 13 & 0.83 & 0.81 & 10.65 & 0.69 & 1.56 & 0.91 & -3.8\\
UGC07151 & 1 & 11 & 3.61 & 0.76 & 11.14 & 3.83 & 3.68 & 0.51 & -3.0\\
UGC07232 & 2 & 4 & 4.60 & 0.52 & 12.57 & 5.36 & 6.33 & 0.47 & +1.0\\
UGC07261 & 2 & 7 & 0.10 & 1.09 & 10.62 & 0.20 & 0.95 & 0.56 & +0.3\\
UGC07323 & 1 & 10 & 1.06 & 0.24 & 12.74 & 1.42 & 0.64 & 0.41 & +5.4\\
UGC07399 & 1 & 10 & 1.60 & 1.25 & 11.13 & 1.70 & 1.82 & 0.64 & -0.1\\
UGC07524 & 1 & 31 & 0.89 & 0.75 & 11.27 & 0.89 & 1.72 & 0.80 & -18.4\\
UGC07559 & 2 & 7 & 1.07 & 0.42 & 10.88 & 3.36 & 2.79 & 0.30 & -5.1\\
UGC07577 & 2 & 9 & 0.34 & 0.24 & 9.16 & 1.92 & 8.29 & 0.23 & -44.0\\
UGC07603 & 1 & 12 & 2.13 & 0.85 & 11.16 & 2.15 & 1.54 & 0.35 & +6.1\\
UGC07608 & 1 & 8 & 0.87 & 0.48 & 12.19 & 1.12 & 0.51 & 0.50 & +4.2\\
UGC07690 & 2 & 7 & 0.84 & 1.28 & 9.98 & 0.95 & 1.51 & 0.60 & -0.5\\
UGC07866 & 2 & 7 & 0.11 & 0.55 & 10.67 & 0.68 & 0.27 & 0.45 & +3.0\\
UGC08286 & 1 & 17 & 2.86 & 1.17 & 10.82 & 2.81 & 2.65 & 1.07 & +2.9\\
UGC08490 & 1 & 30 & 0.13 & 1.33 & 10.61 & 0.14 & 0.33 & 0.86 & +1.0\\
UGC08550 & 1 & 11 & 1.07 & 1.04 & 10.45 & 1.08 & 1.55 & 0.73 & -1.2\\
UGC08699 & 2 & 41 & 0.72 & 0.76 & 12.08 & 0.75 & 0.87 & 0.60 & +0.5\\
UGC08837 & 2 & 8 & 8.24 & 0.16 & 11.37 & 12.53 & 3.07 & 0.18 & +13.5\\
UGC09037 & 2 & 22 & 2.01 & 0.66 & 12.33 & 1.90 & 1.95 & 0.21 & +3.3\\
UGC09133 & 1 & 68 & 7.15 & 0.72 & 12.22 & 7.16 & 7.39 & 1.46 & -21.2\\
UGC09992 & 2 & 5 & 0.04 & 0.83 & 9.47 & 0.17 & 1.24 & 0.50 & +0.8\\
UGC10310 & 1 & 7 & 1.39 & 0.88 & 10.73 & 1.42 & 1.80 & 0.63 & -0.5\\
UGC11455 & 1 & 36 & 5.40 & 0.20 & 13.52 & 5.53 & 4.93 & 0.39 & +11.9\\
UGC11557 & 2 & 12 & 1.58 & 0.36 & 12.33 & 1.68 & 1.29 & 0.46 & +4.4\\
UGC11820 & 1 & 10 & 1.80 & 0.72 & 11.33 & 1.87 & 2.01 & 1.01 & -0.4\\
UGC11914 & 1 & 65 & 0.83 & 0.65 & 13.90 & 0.84 & 1.75 & 0.19 & -49.1\\
UGC12506 & 2 & 31 & 0.18 & 1.30 & 12.02 & 0.24 & 2.16 & 1.14 & -49.1\\
UGC12632 & 1 & 15 & 0.43 & 0.96 & 10.75 & 0.43 & 1.79 & 1.11 & -11.7\\
UGC12732 & 1 & 16 & 0.23 & 0.82 & 11.11 & 0.30 & 0.49 & 1.06 & +1.7\\
UGCA281 & 3 & 7 & 1.31 & 0.49 & 11.73 & 1.80 & 0.47 & 0.37 & +4.6\\
UGCA442 & 1 & 8 & 3.86 & 0.55 & 11.28 & 4.73 & 7.08 & 0.45 & -19.7\\
UGCA444 & 2 & 36 & 0.07 & 0.33 & 11.71 & 0.21 & 0.36 & 0.49 & -2.6\\
\end{tabular}
\end{ruledtabular}
\end{table*}

\bibliography{refs}

\end{document}